\documentclass[pra,aps,superscriptaddress,preprint]{revtex4-2}

\usepackage{graphicx}

\usepackage{amsmath}
\usepackage{bm,CJK}
\usepackage{mathrsfs}
\usepackage{amsfonts}
\usepackage{mathtools}
\usepackage{graphicx}
\usepackage{braket}

\def\be{\begin{equation}}
\def\ee{\end{equation}}
\def\ba{\begin{eqnarray}}
\def\ea{\end{eqnarray}}

\begin{document}

\title{Quantum Algorithm for Approximating Maximum Independent Sets}

\begin{CJK}{UTF8}{gbsn}
\author{Hongye Yu(余泓烨)}
\affiliation{Department of Physics and Astronomy, Stony Brook University, Stony Brook, NY 11794, USA}
\author{Frank Wilczek}
\affiliation{Center for Theoretical Physics, MIT, Cambridge MA 02139 USA}
\affiliation{T. D. Lee Institute, Shanghai Jiao Tong University, Shanghai 200240, China}
\affiliation{Wilczek Quantum Center, School of Physics and Astronomy,
Shanghai Jiao Tong University, Shanghai 200240, China}
\affiliation{Department of Physics, Stockholm University, Stockholm SE-106 91 Sweden}
\affiliation{Department of Physics and Origins Project, Arizona State University, Tempe AZ 25287 USA}
\author{Biao Wu(吴飙)}
\affiliation{International Center for Quantum Materials, School of Physics, Peking University, 100871, Beijing, China}
\affiliation{Wilczek Quantum Center, School of Physics and Astronomy, Shanghai Jiao Tong University, Shanghai 200240, China}
\affiliation{Collaborative Innovation Center of Quantum Matter, Beijing 100871,  China}

\begin{abstract}
We present a quantum algorithm for approximating maximum independent sets of a graph based on quantum non-Abelian adiabatic mixing in the sub-Hilbert space of degenerate ground states, which generates quantum annealing in a secondary Hamiltonian. 
For both sparse and dense random graphs $G$, numerical simulation suggests that our algorithm on average finds an independent set of size close to the maximum size $\alpha(G)$ in low polynomial time.  
The best classical algorithms, by contrast, produce independent sets of size about half of $\alpha(G)$ in polynomial time. 
\end{abstract}
\date{\today}
\maketitle
\end{CJK}
\section{Introduction}

Finding a maximum independent set (MIS) of a graph is an NP-hard problem that appears difficult to solve even approximately. In spite of decades of research, no known classical algorithm produces much better results than  the naive, greedy strategy.

For a graph $G(n,m)$ that contains $n$ vertices and $m$ edges, 
it is known that unless P=NP no polynomial algorithm can find a $ O(n^{1-\epsilon})$-approximate solution in the worst case \cite{johan1999astad}\cite{zuckerman2006linear}, where $\epsilon>0$ is an arbitrary small positive number that is independent of $n$. 
We let $\alpha(G)$ denote the largest size of independent sets for a given graph $G$. The aforementioned statement means that 
the size of the best approximate MIS found by a polynomial algorithm is $\sim \alpha(G)/n^{1-\epsilon}$. This is not an 
impressive result when you notice that $1\le \alpha(G)\le n$.
Average case performance, for both sparse and dense graphs, is not much better.  Consider for example the class of
Erd\H{o}s-R\'{e}nyi random graphs, denoted $G(n,\mathcal{P})$, where $\mathcal{P}$ is the probability to generate 
an edge between any pair of vertices.  
Erd\H{o}s-R\'{e}nyi graphs $G(n,\mathcal{P})$ are dense at $\mathcal{P}=1/2$, as their edge numbers are proportional to $n^2$. For them, the MIS size is typically $\alpha(G(n,1/2))\sim 2 \log_2 n$~\cite{matula1976largest}.
But no classical algorithm is known or suspected to produce in polynomial time, with non-vanishing probability, an independent set of size $(1 + \epsilon) \log_2 n$ for any fixed $\epsilon> 0$ , neither analytically nor through numerical evidence \cite{coja2015independent}. 
It is common to take $d = 2m/n$ to define sparse random graphs $G(n,m)$ parametrically. One finds that for sparse graphs with $d\gg 1$~\cite{frieze1990independence} 
\be
\alpha(G(n,m))\sim 2n \frac{\ln d}{d}\,.
\ee 
Here too, no classical algorithm is known or suspected to perform well - specifically, to find an independent set of size $(1 + \epsilon)n \frac{\ln d}{d}$ 
in polynomial time with non-vanishing probability.

Here we introduce a quantum algorithm which appears, in extensive numerical evidence, to perform much better.  It builds on the 
the quantum algorithm for independent sets we proposed in Ref. \cite{wu2020quantum}, but adds a major new ingredient.  
Numerical experiments indicate that our quantum algorithm typically produces an independent set of size almost $\alpha(G)$ in low polynomial time, 
for both sparse and dense random graphs in the average-case scenario, where we average over both the final 
quantum measurement and many randomly generated graphs.

\section{Quantum adiabatic evolution in the solution-subspace}

Our approach for approximating MIS builds on a quantum algorithm for independent sets~\cite{wu2020quantum,Yanglin}.  To fix notation and to make this work self-contained, we briefly recall the earlier algorithm here.
For a given non-empty graph $G$, we construct  a corresponding spin-system with the following Hamiltonian~\cite{wu2020quantum}
\begin{equation}
\label{spinH}
H_{0}=\Delta\sum_{\langle ij\rangle}(\hat{\sigma}^z_{i}+\hat{\sigma}^z_{j}+\hat{\sigma}^z_{i}\hat{\sigma}^z_{j})\,,
\end{equation}
where the summation $\langle ij\rangle$ is over all edges in the graph. Spin $j$ being up should be interpreted as inclusion of site $j$ in the candidate set, and the terms in the Hamiltonian as imposing a penalty for connection between included sites.  Two key features of this Hamiltonian are:
\begin{enumerate}
    \item The independent sets of a graph $G$ are in one-to-one correspondence with the ground states of $H_0$. 
    \item There is an energy gap $4\Delta$ between the ground states 
and the first excited states, independent of $n$ and $m$.
\end{enumerate} 
These features allow us to explore the space of independent sets through non-abelian adiabatic evolution.

\def\ur{u_{\vec{r}}}
\def\dr{d_{\vec{r}}}

We consider acting uniformly upon all the spins with the rotation matrix 
\begin{equation}
V_j=
\begin{pmatrix}\cos\frac{\theta}{2} & e^{-i\varphi}\sin\frac{\theta}{2}\\ 
	e^{i\varphi}\sin\frac{\theta}{2}&-\cos\frac{\theta}{2} 
\end{pmatrix} 
=V_j^{-1}\,.
\label{matrix}
\end{equation}
$V_j$ represents rotation through $\pi$ around the axis $(\sin\frac{\theta}{2} \cos\varphi, \sin\frac{\theta}{2} \sin\varphi, \cos\frac{\theta}{2})$ and takes the unit vector $(0, 0, 1)$ to $\vec r \equiv(\sin\theta \cos\varphi, \sin\theta \sin \varphi, \cos\theta)$. Note that such mapping from the $SO(3)$ rotation to $V_j$ is not unique and $V_j$ does not have unit determinant, but it includes a convenient overall phase factor.


If $\ket{u}_j$ and $\ket{d}_j$ are eigenstates 
of $\hat{\sigma}_j^z$,  that is, $\hat{\sigma}_j^z\ket{u}_j=\ket{u}_j$ and  $\hat{\sigma}_j^z\ket{d}_j=-\ket{d}_j$, 
then the eigenstates of $\hat{\tau}^z_j \equiv V_j \hat{\sigma}_j^z V_j^{-1}$ are 
\begin{eqnarray}
\label{npm}
\ket{\ur}_j&=&\cos\frac{\theta}{2}\ket{u}_j+\sin\frac{\theta}{2}e^{i\varphi}\ket{d}_j~,\\
\ket{\dr}_j&=&\sin\frac{\theta}{2}\ket{u}_j-\cos\frac{\theta}{2}e^{i\varphi}\ket{d}_j~.
\end{eqnarray}
Upon acting with $U=V_1\otimes V_2\otimes \cdots \otimes V_n$, 
we rotate all the spins and find a new Hamiltonian
\begin{equation}
H_\tau=UH_{0}U^{-1}=\Delta\sum_{\langle ij\rangle}(\hat{\tau}^z_{i}+\hat{\tau}^z_{j}+\hat{\tau}^z_{i}\hat{\tau}^z_{j})~.
\end{equation}
Of course $U$ need not be implemented through physical rotation of the computing apparatus; it can be simulated using parallel operation of simple, one-bit gates.

$H_\tau$ has the same set of eigenvalues as $H_0$.  Its eigenstates of $H_\tau$ are
obtained  by rotating those of  $H_0$, in the form

\begin{equation}
\ket{E_\mu(\theta,\varphi)} =\ket{s_1}\otimes \ket{s_2}\otimes \cdots\otimes\ket{s_j}\otimes\cdots \otimes
\ket{s_n}\nonumber
\end{equation}
where $\mu$ is a string of $\{\pm 1\}^n$ and $\ket{s_j} = \ket{\ur} \ {\rm or} \ \ket{\dr}$ for $\mu_j = 1 \  {\rm or} \ -1$.

The quantum algorithm in Ref. \cite{wu2020quantum} starts the spin system in the state $E_\mu$ with 
$\mu=\{-1,-1,\cdots,-1\}$, which is one of many ground states of $H_{0}$. Then all spins are rotated in the same way, 
by slowly changing $\vec{r}$. 
The system evolves, to exponential accuracy in the slowness parameter, within the sub-Hilbert space spanned by the ground 
states of $H_\tau$.  But the evolution within that space is nontrivial due to the non-Abelian geometric phase~\cite{WZ}, and when $\vec{r}$ is rotated back to the $z$-direction, 
upon measurement one obtains with high probability a non-trivial independent set~\cite{wu2020quantum}.

The evolution within the sub-Hilbert space 
of the ground states is given by~\cite{WZ}
\be
\ket{\psi(t)}=P \exp \left(i\int_{0}^{t} A(t') dt' \right)\ket{\psi(0)}\,,
\label{gauge}
\ee
where $P$ stands for time ordering and $A$ is the hermitian nonabelian gauge matrix. 
The off-diagonal terms of  the gauge matrix $A$ are non-zero only  when they connect states labelled by strings $\mu, \nu$ separated by Hamming distance $|\mu-\nu|=1$.  In that case we have
\begin{equation}
\label{Aij}
A_{\mu, \nu}(\theta)=i\left\langle E_{\mu}\left|\partial_{t}\right| E_{\nu}\right\rangle= i\left\langle \ur\left|\partial_{t}
\right| \dr\right\rangle=\frac{\sin \theta}{2} \frac{d \varphi}{d t}+\frac{i}{2}\text{sgn}(\mu-\nu)\frac{d \theta}{d t}\,,
\end{equation}
where $\text{sgn}(\mu-\nu)$ is a sign function, depending on the sign of first non-zero element of $\mu-\nu$. And $\mu-\nu$ is defined as element-wise subtractions ( e.g.,\ $\text{sgn}(\{1,1,-1\}-\{-1,1,1\})=\text{sgn}(\{2,0,-2\})=+1$).
The diagonal terms of $A$ are
\begin{equation}
\label{Aii}
\begin{aligned} A_{\mu, \mu}(\theta) &=i\left\langle E_{\mu}\left|\partial_{t}\right| E_{\mu}\right\rangle 
=-\left\{n_+ \sin ^{2} \frac{\theta}{2}+(n-n_+) \cos ^{2} \frac{\theta}{2}\right\} \frac{d \varphi}{d t}\,, \end{aligned}
\end{equation}
where $n_+$ is the number of plus signs in $\mu$.

Eq. (\ref{gauge}) indicates that that the gauge matrix $A$ can 
be regarded as an emergent Hamiltonian for the spin system, generating unitary evolution within the  eigenspaces of the original  Hamiltonians $H_0$.  We call this the secondary Hamiltonian.  In Ref.\cite{wu2020quantum}, 
we took $\theta$ to be fixed and let $\varphi$ vary slowly.  This gives rise to a
time-independent secondary Hamiltonian $A(\theta)$.  In this work we change both $\varphi$ and $\theta$ slowly, under 
the condition $d\theta/dt\ll d\varphi/dt$. This generalization brings in profoundly different dynamics.
In this case, $A(\theta)$ becomes a time-dependent secondary Hamiltonian with the parameter $\theta$ changing slowly.  Remarkably, 
the empty-set solution $\mu=\{-1,-1,\cdots,-1\}$ is the ground state of $A(0)$, but the maximum independent set (MIS), 
which has largest number of vertices $n_+$, is the ground state of $A(\pi)$.

According to the adiabatic theorem, sufficiently slow evolution of the secondary Hamiltonian will keep us within the ground state manifold. This means that 
if we change $\theta$ slowly enough, and evolve from $\theta = 0$ to $\theta = \pi$, we
will evolve to the state representing the maximum independent set when $\theta=\pi$. (Note that at the end we must reverse the spin directions, 
e.g. turning $\{-1,-1,+1\}$ into $\{+1,+1,-1\}$, as the system ends along the $-z$ direction ($\theta=\pi$).)This is a quantum adiabatic algorithm for MIS.  Its  time complexity is determined by
the energy gap of $A(\theta)$~\cite{Farhi2000}. In a worst case scenario 
the energy gap of $A(\theta)$ can be exponentially small, as we will shortly exemplify.
However, our numerical results show that more typically, in interesting cases we 
get independents set whose size is very  close to $\alpha(G)$.

\section{Two special graphs}
To illustrate possible behavior of the minimum energy gap of $A(\theta)$, let us consider two special graphs. 
The first graph is the one that has no edges. 
In this case, all combinations of vertices are independent sets and the gauge matrix $A(\theta)$ acts on the whole $2^n$-dimension Hilbert space. 
Denote $A(\theta)$ for no-edge graphs as $\tilde{A}$. It can be re-written as
\begin{equation}
	\tilde{A}(\theta)=\frac{\sin \theta}{2}\frac{d \varphi}{d t}\sum_{j=1}^n\tilde{\sigma}_j^x+
	\frac{\cos\theta }{2}\frac{d \varphi}{d t}\sum_{j=1}^n\tilde{\sigma}_j^z+
	\frac{1}{2}\frac{d \theta}{d t}\sum_{j=1}^n\tilde{\sigma}_j^y+(\cos\theta-n\cos^2\frac{\theta}{2})I
\end{equation}
where $I$ is the $2^n\times 2^n$ identity matrix and  contributes only a global phase factor during the evolution. 
Note that these $\tilde{\sigma}_j^x,\tilde{\sigma}_j^y,\tilde{\sigma}_j^z$ are  not the spin operators $\sigma_j^z$ in $H_0$, and they 
are used just to  put $\tilde{A}(\theta)$ in a concise form.  If  $d \theta/d t$ is much smaller 
than $d \varphi/d t$, then we can omit the third term of $\tilde{A}$ and have
\begin{equation}
\tilde{A}(\theta)\approx\frac{\sin \theta}{2}\frac{d \varphi}{d t}\sum_j\tilde{\sigma}_j^x+\frac{\cos\theta }{2}\frac{d \varphi}{d t}
\sum_j\tilde{\sigma}_j^z\,.
\end{equation}
This is effectively a Hamiltonian for $n$ identical non-interacting spins in the same magnetic field.  
Apparently, $\tilde{A}(\theta)$  has a constant gap between the ground state and the first excited state. 
When we let $\theta$  evolve slowly from 0 to $\pi$ for a fixed period of time,  the system no matter how large will  
evolve from the initial ground state at $\theta=0$ to the ground state at $\theta=\pi$.  
This is consistent with the original Hamiltonian in Eq.(\ref{spinH}). For the graph with no edges, the Hamiltonian $H_0$ is zero. 
This means that there is no evolution; the system stays in the state $\{-1,-1,\cdots,-1\}$. Upon reversing the direction of the spins, we get the MIS  $\{1,1,\cdots,1\}$. 

\begin{figure}[h]
	\centering
	\includegraphics[width=0.9\textwidth]{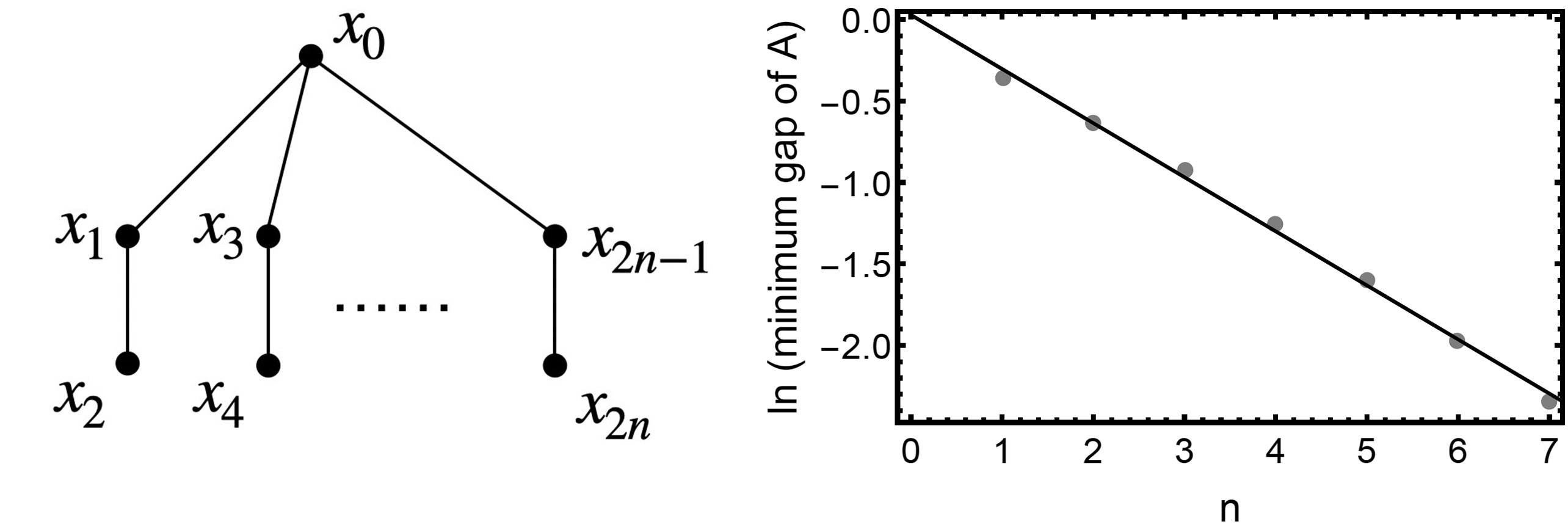}
	\caption{(Left) A special type of graphs that has $2n$ edges and $2n+1$ vertices. 
	Note that it has a unique maximum independent set $\{x_0, x_2,...,x_{2n}\}$. 
	(Right) the minimum energy gap of $A$ for these graphs as a function of $n$. The fitting line is given by ln(gap)=$0.0286 -0.332 n$. }
	\label{ce1}
\end{figure}

The second special graph $S_n$ is shown in Fig.\ref{ce1}, which has $2n+1$ vertices and $2n$ edges. The graph has $2^n$ maximal independent sets, 
and only one of them is the MIS.   For each $n$, we compute numerically the energy gaps of $A(\theta)$ for $0\le\theta\le\pi$ and find the minimum. 
The results are plotted in Fig.\ref{ce1}, which shows that the minimum energy gaps of $A(\theta)$ 
for these graphs decrease exponentially with $n$.  

It is clear from these two special types of graphs that there is no universal behavior for the minimum energy gap of the gauge matrix $A(\theta)$.

\section{Quantum algorithm for approximately maximal independent set}
Our quantum algorithm for finding an approximately maximal independent set runs as follows:
\begin{quote}
1. Construct the Hamiltonian $H_0$ according to a given graph $G$ and prepare the system at the state $\{-1,-1, \cdots,-1\}$.\\
2. Set $\vec{r}(\theta,\varphi)$ initially along the $z$ axis and slowly change $H_{\tau}$ according to 
$\theta=\omega_\theta t, \varphi=\omega_\varphi t$ 
with $\omega_\varphi$ being some constant and $\omega_\theta=\pi\omega_\varphi/T$. $T=n^\gamma$ is the total run time.\\
3. Stop the system at $\theta=\pi$ and make a measurement along the $z$ axis. \\
4. Reverse the direction of the spins, e.g. changing $\{-1,-1,\cdots, -1\}$ into $\{1,1,\cdots,1\}$, to achieve the $\mu$ for the candidate answer. 
\end{quote}
Since the energy gap of the secondary Hamiltonian $A$ can be exponentially small, run times $T=n^\gamma$ which scale polynomially do not guarantee that the system will stay in the ground state of $A$. However, the system will stay mostly within the manifold of states whose energy is close to the ground state, i.e., approximately maximum states, if the evolution is slow enough.
As a result, at the end of computation, we might expect to find a good approximately maximum states. We have explored this hypothesis numerically, with excellent results in generic cases, as we will now discuss.  As the adiabatic condition for $H_{\tau}$ can be
satisfied (see supplemental online materials), our numerical simulation is done with $A$ so that we can compute for larger graphs. 

\begin{figure}[h]
	\centering
	\includegraphics[width=0.9\textwidth]{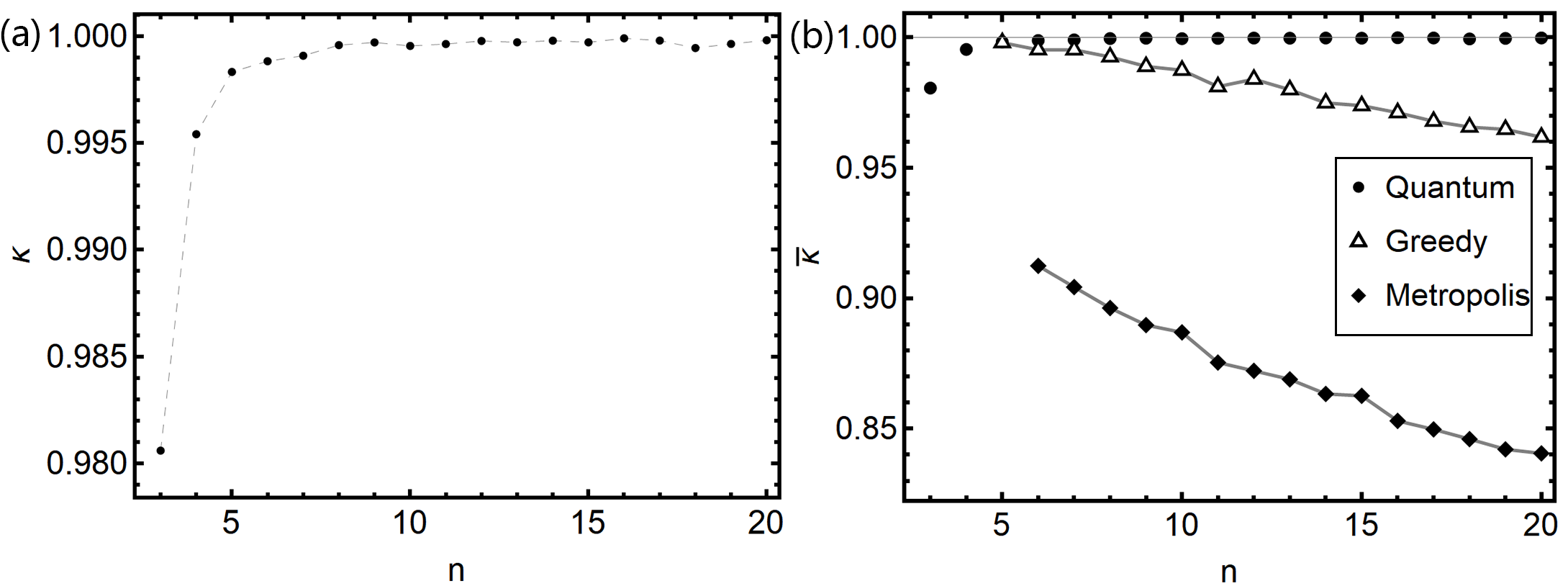}
	\caption{The average $\kappa$ (or $\bar{\kappa}$) as a function of $n$ for Erd\H{o}s-R\'{e}nyi random graphs $G(n,1/2)$ with three different algorithms. (a) The 
		results of our quantum algorithm. We set $T=n^2$, $\omega_\varphi=1$, $\omega_\theta=\pi/T$ and run over 1000 Erd\H{o}s-R\'{e}nyi random graphs $G(n,1/2)$. The variance of $\kappa$ is around $10^{-6}$. The calculation is done
		with $A$. (b) The results of the Greedy algorithm and the Metropolis algorithm in comparison with our quantum results. For the Greedy algorithm,  the calculation runs 1000 times over one graph to get $\bar{N}$, and then runs over 1000 random graphs to get $\bar{\kappa}$. The variance of $\bar{\kappa}$ is around $10^{-4}$. For the Metropolis algorithm, we set the iteration time $T=n^2$. 
		The calculation runs 1000 times over one graph to get $\bar{N}$, and then runs over 1000 random graphs to get $\bar{\kappa}$. 
		The variance of $\bar{\kappa}$ is around $10^{-4}$. The  lines in the figure are guide for the eye. }
	\label{tn2}
\end{figure}

If the final quantum state is $\ket{\psi_f}=\sum_\ell a_\ell\ket{E_\ell}$ (after the reverse of the spin direction), 
we define the averaged size $\bar{N}$ of the independent sets as
\begin{equation}
\bar{N}=\sum_\ell |a_\ell|^2N_\ell
\end{equation}
where $N_\ell$ is the size of the $\ell$th independent set $\ket{E_\ell}$. In the quantum mechanical formalism, this represents the average value of a single measurement.  We are interested in the ratio  $\kappa =\bar{N}/\alpha(G)$.
Our numerical results, displayed in  Figure \ref{tn2}(a), show that for Erd\H{o}s-R\'{e}nyi random graph $G(n,1/2)$, if we set $T\sim n^2$, 
the average of $\kappa$ will increase to almost 1 when $n$ increases.  This is compared to the results obtained using the 
classical greedy and Metropolis algorithms~\cite{doi:10.1002/rsa.3240030402} (see supplemental online material for details of the two algorithms). 
We run the classical algorithms several times on each graph to get $\bar{N}$ and ratio $\kappa=\bar{N}/\alpha(G)$, then run the process over multiple random graphs to find the double average $\bar{\kappa}$. Our numerical results in  Fig.\ref{tn2}(b) show that even for small graphs, the ratio $\bar{\kappa}$ in the two classical algorithms 
is not as close to 1 as the one with our quantum algorithm. More importantly, 
the classical ratio $\bar{\kappa}$ tends to decrease when $n$ gets larger. This is consistent with the well known result
that the best classical polynomial algorithm face grave difficulty in
pushing the ratio larger than $1/2$ when $n$ goes to infinity \cite{coja2015independent} (see below).

\begin{figure}[h]
	\centering
	\includegraphics[width=0.9\textwidth]{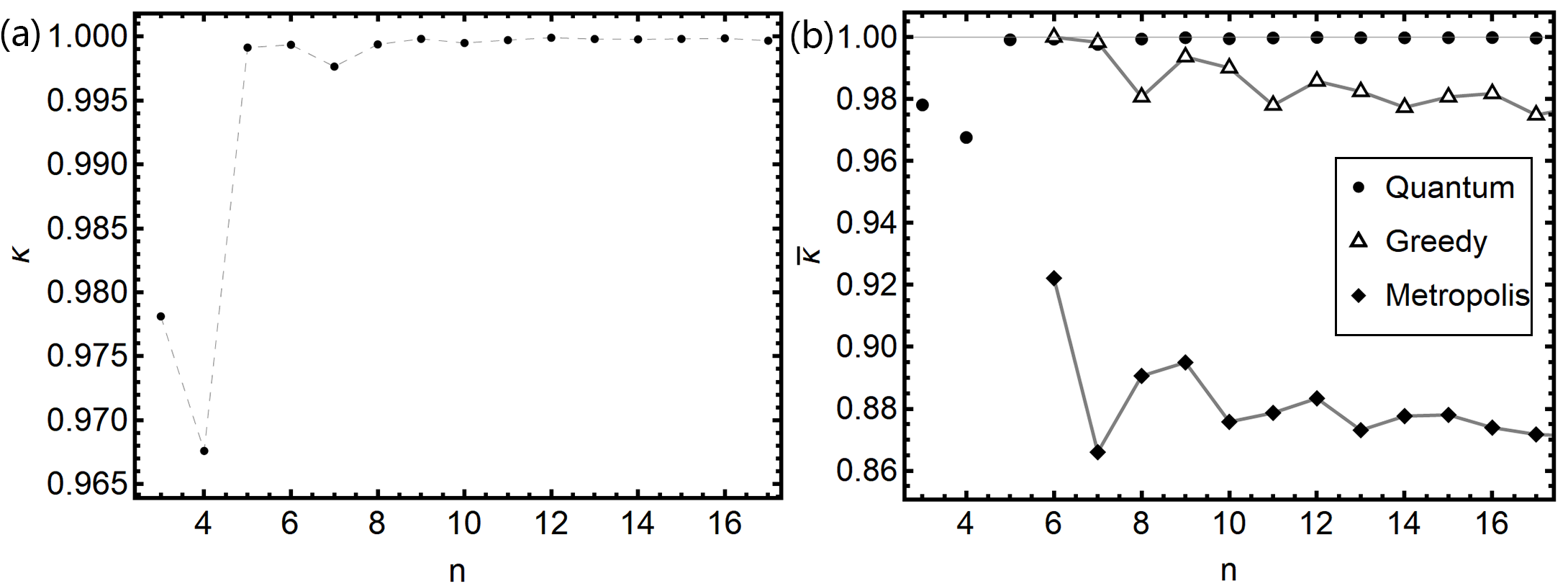}
	\caption{The average $\kappa$ (or $\bar{\kappa}$) as a function of $n$ for random graphs $G(n,m)$ with $m=\lfloor n \ln n\rfloor$ via three different methods.  (a) The 
	results of our quantum algorithm. We set $T=n^2$, $\omega_\varphi=1$, $\omega_\theta=\pi/T$ and run over 1000  random graphs $G(n,\lfloor n \ln n\rfloor)$. The variance of $\kappa$ is around $10^{-5}$.  The calculation is done with $A$. (b) The results of the Greedy algorithm and the Metropolis algorithm in comparison with our quantum results. For the Greedy algorithm,  the calculation runs 1000 times over one graph to get $\bar{N}$, and then runs over 1000 random graphs to get $\bar{\kappa}$. The variance of $\bar{\kappa}$ is around $10^{-4}$. For the Metropolis algorithm, we set the iteration time $T=n^2$. 
	The calculation runs 1000 times over one graph to get $\bar{N}$, and then runs over 1000 random graphs to get $\bar{\kappa}$. 
	The variance of $\bar{\kappa}$ is around $10^{-4}$. The lines in the figure are guide for the eye.}
	\label{tn}
\end{figure}

For sparse graphs with edge number $m=\lfloor n \ln n\rfloor$ the results are similar, as shown in Figure \ref{tn}.

These numerical results indicate 
that our quantum algorithm can find an independent set of size $(1-\epsilon)\alpha(G)$ in run times $T\sim n^2$. 
We also tried $T\sim n$. In this case the average radio $\kappa$ decreases when $n$ increases. 
These numerical results suggest that our quantum algorithm is of time complexity of $O(n^2)$. 
Note that in Ref.~\cite{Regev}, it was shown that the run time required in a quantum adiabatic algorithm can increase polynomially with 
the system size even when the energy gap is constant. Our numerical results (see the supplemental online material) show
that the adiabaticity for $H_\tau$ is ensured with $T\sim n^2$.

\section{Diffusion and annealing in solution trees}

In this section, we review a theoretical picture that clarifies the challenge of finding approximate maximum independent sets, and offer an heuristic explanation for the enhanced performance of our quantum algorithm, relative to classical ones.

\begin{figure}[h]
	\centering
	\includegraphics[width=0.45\textwidth]{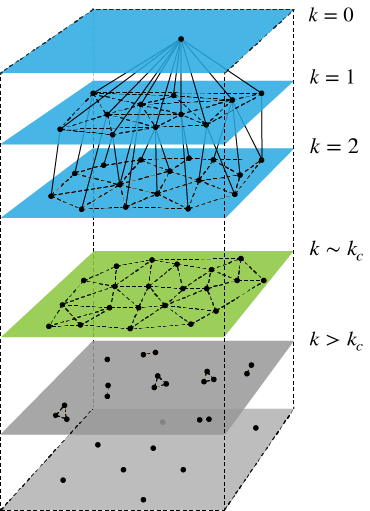}
	\caption{ (color online)The tree of independent sets of a graph $G$. Each point represents an independent set; the one at the top represents the empty set. 
	The tree is layered: the independent sets $S_k(G)$ in each layer has the same size $k$. 
	If the Hamming distance between an independent set of size $k$ and an independent set of $k+1$ is one, they are connected by a solid line.
	Each point in the layer of $k+1$ must be connected by a solid line with a point in the layer of $k$. 
	For clarity, we only draw the solid lines between $k=0$ and $k=1$ and between $k=1$ and $k=2$. For independent sets of the same size, 
	they are connected by dashed lines if the Hamming distance between them does not scale up with $n$. 
	Before a critical size $k_c$, 
	the tree is well connected by dashed lines in each layer. When the size is over $k_c$, the layers shatter with the independent sets divided into small groups, between each pair of which the Hamming distance is proportional to $n$. At the same time,  
	the group size decreases exponentially with $n$. }
	\label{shatter}
\end{figure}

For sparse graphs $G(n,m)$,  Coja-Oghlan and Efthymiou showed in Ref. \cite{coja2015independent} 
that the difficulty is related to the structure of the space of independent sets, which shatters severely when their size $k$ is large enough.  
Thus, the classical Metropolis process has exponentially large mixing times. The graphs considered in Ref. \cite{coja2015independent} 
have $d=2m/n\gg1$. 
For these graphs, the size of the maximum independent set is $\alpha\sim (2-\epsilon_d)n\frac{\ln d}{d}$ with high probability. 
Let $S_k(G)$ denote all the independent sets of size $k$. ``$S_k(G)$ shatters severely" in the precise sense that $S_k(G)$ can be divided into 
many groups such that the Hamming distance between each pair of groups is  proportional to $n$, while the number of 
independent sets in each group decreases exponentially with $n$~\cite{coja2015independent} (see Figure \ref{shatter}).  
It is found that  $S_k(G)$ shatters for $(1+\epsilon_d)n\frac{\ln d}{d}<k<\alpha$. This means that 
searches for the maximum independent set, based on building up through consideration of changes in small numbers of entries will 
get stuck at sizes around $n\frac{\ln d}{d}$. This is the essential reason why 
polynomial classical algorithms have difficulty finding independent sets of size $k>(1+\epsilon_d)n\frac{\ln d}{d}$. 

Quantum evolution, by allowing superpositions, can enable more efficient exploration of a shattered solution landscape.  All the candidates appear as components in the wave function. In our context, different independent sets are assigned different energies according to the secondary Hamiltonian. During slow evolution, we can expect the system - which starts cold, and plausibly remains so, to approach a quasi-thermal equilibrium state, favoring larger overlaps with lower energy eigenstates.  Since low energies correspond, at the conclusion of the evolution, to approximate maximum independent sets, with high probability they will appear as the result of the final measurement. 
This argument is far from rigorous, but it makes the striking numerical results presented above appear less mysterious.   

It has been rigorously established that quantum diffusion 
can hold advantages over classical random walk for a special types of decision trees\cite{QTree}. In the future, it will be important to investigate 
further why and in what circumstances quantum diffusion is more effective than its classical counterpart.  




\section{Conclusion}
We have proposed a quantum algorithm for approximating the maximum independent set of a graph $G(n,m)$ by exploiting non-Abelian adiabatic mixing
in the sub-Hilbert space of solutions and adiabatic evolution in the secondary Hamiltonian it generates. Our numerical experiments indicate that for both sparse and dense graphs on average we obtain an independent set of almost maximum size $\alpha(G)$ size  in 
the evolution time $T\sim n^2$ with a single measurement.  

While our numerical results are encouraging, they are limited to relatively small systems. Due
to the exponential complexity of simulating qubit systems classically, we only calculated systems containing up to 20 qubits.  We made a heuristic argument that makes a robust quantum advantage, extending to large $n$, seem plausible, but this question deserves much further attention.  

\acknowledgments
 FW is supported in part by the U.S. Department of Energy under grant DE-SC0012567, by the European Research Council under grant 742104, and by the Swedish Research Council under contract 335-2014-7424.  BW is supported by the The National Key R\&D Program of China (Grants No.~2017YFA0303302, No.~2018YFA0305602), National Natural Science Foundation of China (Grant No. 11921005), and 
Shanghai Municipal Science and Technology Major Project (Grant No.2019SHZDZX01).


\end{document}